# The Infrared Imaging Spectrograph (IRIS) for TMT: the ADC optical design

Andrew C. Phillips*[a], Ryuji Suzuki[b], James E. Larkin[c], Anna M. Moore[d], Yutaka Hayano[b], Toshihiro Tsuzuki[b], Shelley A. Wright[e]

[a]University of California Observatories, UC Santa Cruz; [b]National Astronomical Observatory of Japan; [c]University of California, Los Angeles; [d]Caltech Optical Observatories; [e]University of California, San Diego

## ABSTRACT

We present the current optical design for the IRIS Atmospheric Dispersion Corrector (ADC). The ADC is designed for residual dispersions less than ~1 mas across a given passband at elevations of 25 degrees. Since the last report, the area of the IRIS Imager has increased by a factor of four, and the pupil size has increased from 75 to 90mm, both of which contribute to challenges with the design. Several considerations have led to the current design: residual dispersion, amount of introduced distortion, glass transmission, glass availability, and pupil displacement. In particular, it was found that there are significant distortions that appear (two different components) that can lead to image blur over long exposures. Also, pupil displacement increases the wave front error at the imager focus. We discuss these considerations, discuss the compromises, and present the final design choice and expected performance.

**Keywords:** ADC, Atmospheric Dispersion Corrector, Atmospheric Dispersion Compensator, IRIS, InfraRed Imaging Spectrograph, Imager, Spectrograph

## 1. INTRODUCTION

The InfraRed Imaging Spectrograph (IRIS)[1] is a first light instrument on the Thirty-Meter Telescope (TMT) now in its construction phase. IRIS is fed by the Narrow Field InfraRed Adaptive Optics System (NFIRAOS)[2] and is designed to produce diffraction-limited images across its field of view, a square 35 arc seconds on each side. The wavelength range is $0.84 < \lambda < 2.45$-μm. IRIS has both imaging and integral-field unit (IFU) spectroscopic modes.

At limiting elevations, atmospheric dispersion can be an order of magnitude larger than the diffraction-limited images even across a single IR passband. IRIS science cases requires the atmospheric dispersion to be corrected within each passband to better than 1 milli-arcsecond (mas). The goal is to be able to achieve astrometric precision on the order of 10 micro-arcseconds (μas). The design of the Atmospheric Dispersion Corrector (ADC) has been challenging. In addition to finding glasses suitable for the correction required, several other problems such as distortion have become apparent that have needed to be dealt with.

Phillips et al.[3] presented a conceptual design for the IRIS ADC, and gave the reasons for selecting crossed-Amici prisms in collimated light as the basic design. At that time, IRIS had independent imaging and spectroscopic paths, both with independent ADCs. It was decided (in part because of problems with large wedge angles on the spectrograph prisms) that the spectroscopic path would be downstream from the imaging path, fed by a pick-off mirror near the focal plane of the detector. This means that a single ADC now corrects for both the imaging and spectroscopic modes. Another major design change in IRIS during this intervening time is the increase of the field of view; the design now calls for four rather than one H4RG detectors, with the imaging area doubled on each side. Yet another major change since the conceptual design is that an actual sire, MK-13 on Mauna Kea was selected, enabling us to design for a specific altitude, although currently the final TMT site is again in question. Lower altitude sites would have a significant impact on the ADC design and performance.

This paper describes the current ADC design for IRIS, and the reasons behind design decisions. Section 2 discusses the search for suitable glass pairs. Section 3 describes briefly the concept of a "global" ADC in NFIRAOS, an idea that was eventually deemed impractical. Section 4 discusses ADC-induced distortion in some detail. Finally, Section 5 discusses the down-select to a specific glass pair, and its actual design parameters and expected performance. For the sake of clarity in the discussion, we define various angles and spaces in Figure 1.

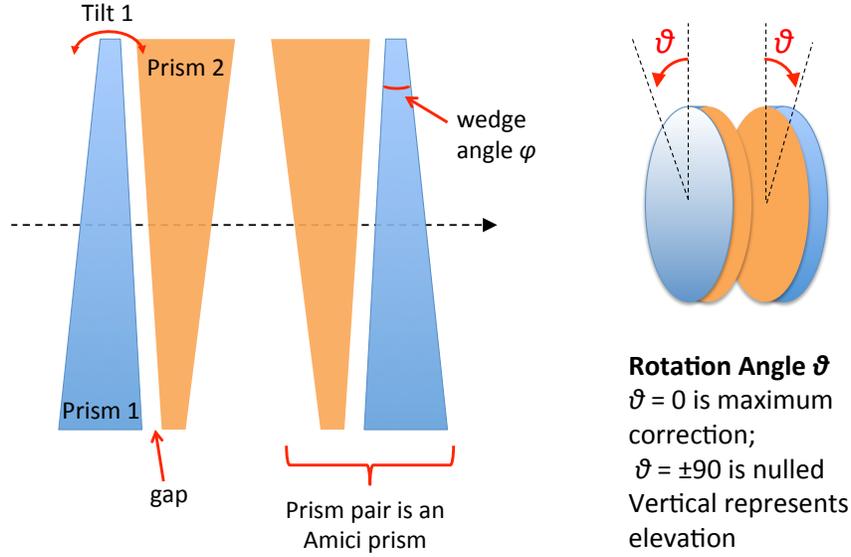

Figure 1. Two schematic views of the ADC, with various terms defined.

## 2. GLASS PAIRS

After the conceptual design, we decided to do an extensive search of additional glass pairs. We identified all glasses in the Ohara catalogue that have reasonable (>0.82) 1-cm transmission at 2.4 µm. This included some glasses that had been recently added to the catalogue. Some other materials (e.g. spinel, $CaF_2$) were also included. Glass indices were adjusted to the relevant temperature and pressure (77K and 0 for IRIS; -30C and 1 atm for NFIROAS). These were sorted into high-dispersion and low-dispersion glasses using a NIR "Abbé diagram." We then solved the best linear combination of 831 glass/material pairs to minimize the residual dispersion over the ranges $0.84 < \lambda < 2.45$ µm and $0.58 < \lambda < 2.45$ µm (the latter to consider a solution that included the Na I D wavelength for Laser Guide Star applications). Those with low rms residual dispersion across these ranges, particularly the IRIS range, were selected for further consideration. Table 1 lists the pairs with smallest residuals, along with wedge angles α and slope across the pupil diameter.

We note that the glass pair selected in the conceptual design, S-NPH2+spinel, occurs at the end of this table. This is in part to the redefinition of the IRIS range. However, spinel had already been rejected as a material due to difficulty in obtaining optical-quality blanks and polishing limitations.[4]

## 3. THE GLOBAL ADC OPTION

There are distinct advantages to putting a large ADC in the beam inside of NFIRAOS. For natural guide stars, a correction for atmospheric dispersion would deliver sharper images at the wave front sensor (WFS) and hence better performance of the adaptive optics. Furthermore, such an ADC would also provide dispersion correction to the IRIS On-Instrument Wave Front Sensors (OIWFS) used for tip-tilt/distortion corrections to the images supplied by NFIRAOS, as well as correcting all IRIS science-mode paths. ADC-introduced distortions are also minimized.

The crossed-Amici design needs collimated space. We identified a suitable location within NFIRAOS near the pupil between DM12 and OAP2. There were issues concerning mechanical packaging, but these were set aside until the optical performance could be seen. ZEMAX was used to model the ADC performance through to the focal plane at the exit to NFIRAOS.

It was found that an ADC could be added to NFIRAOS, but several concerns arose. First, the dispersion correction had to cover a longer wavelength range, from the Na I D wavelength λ593 nm, through the natural guide star range (600 < λ < 800 nm), and finally through the entire IRIS range. It turns out that the glass pair S-LAH71+S-FPL51 could do this adequately. However, the fact that there was only one ADC meant that, for instance, it would be impossible to tune the ADC for individual passbands for IRIS science without significantly impacting dispersion correction in NFIRAOS and the OIWFS. From this point of view, separate ADCs for the natural guide star channel in NFIRAOS and in the OIWFS make more sense.

In the end, the global ADC option was rejected, simply because only one suitable glass pair, S-BAH28+S-BSM2, was found to be available in the size required (~350 mm diameter). This pair has only marginal transmission at 2.4 μm; with the thickness required for these large prisms, the transmission at λ > 2 μm was deemed unacceptably low. Considering the impact on IRIS science, the global ADC option was rejected.

Table 1. Glass pairs, wedge angles, slopes, and residual dispersion across the entire IRIS wavelength range.

| glnam1 | α1 | slope1 (mm) | glnam2 | α2 | slope2 (mm) | rms (mas) |
|---|---|---|---|---|---|---|
| S-TIM39 | -1.8 | -11.1 | S-BSM2 | 2.0 | 12.0 | 0.82 |
| S-LAH79 | -0.7 | -4.1 | S-FPM3 | 1.2 | 7.5 | 0.84 |
| S-TIM28 | -1.6 | -9.7 | S-BSM18 | 1.7 | 10.3 | 1.06 |
| S-FTM16 | -2.3 | -14.1 | S-BAL12 | 2.5 | 15.3 | 1.10 |
| S-NBH56 | -0.8 | -4.9 | S-BSM2 | 1.1 | 6.8 | 1.14 |
| S-FTM16 | -2.3 | -14.1 | S-BAL42 | 2.3 | 14.2 | 1.21 |
| S-NPH5 | -0.7 | -4.5 | S-BSM9 | 1.0 | 6.1 | 1.28 |
| S-NPH1 | -0.8 | -4.7 | S-BAL14 | 1.1 | 6.5 | 1.28 |
| S-TIM28 | -1.9 | -11.4 | S-BSM28 | 2.1 | 12.5 | 1.29 |
| S-FTM16 | -2.5 | -15.1 | S-BAL14 | 2.5 | 15.5 | 1.53 |
| S-NPH3 | -0.4 | -2.7 | S-BAL14 | 0.7 | 4.4 | 1.53 |
| S-BAH28 | -2.4 | -14.7 | S-BSM2 | 2.8 | 17.4 | 1.55 |
| S-TIM35 | -1.5 | -9.0 | S-BSM9 | 1.7 | 10.1 | 1.57 |
| S-NPH53 | -0.8 | -4.9 | S-BSM18 | 1.0 | 6.4 | 1.65 |
| S-TIM3 | -2.7 | -16.7 | S-BAL14 | 2.9 | 17.8 | 1.74 |
| S-NPH1 | -0.8 | -4.6 | S-BAL12 | 1.1 | 6.7 | 1.83 |
| S-NPH3 | -0.4 | -2.7 | S-BAL12 | 0.8 | 4.6 | 1.87 |
| S-NPH4 | -0.6 | -3.5 | spinel | 0.7 | 4.3 | 1.91 |
| S-NPH53 | -0.9 | -5.4 | S-BSM28 | 1.2 | 7.2 | 1.92 |
| S-NPH3 | -0.4 | -2.7 | S-BAL42 | 0.7 | 4.2 | 1.93 |
| S-NPH1 | -0.8 | -4.6 | S-BAL42 | 1.0 | 6.2 | 1.98 |
| S-TIH4 | -1.0 | -6.0 | S-PHM52 | 1.2 | 7.1 | 2.04 |
| S-NBH56 | -1.0 | -6.3 | S-BAM4 | 1.4 | 8.7 | 2.10 |
| S-TIH11 | -0.9 | -5.3 | S-PHM52 | 1.1 | 6.6 | 2.11 |
| S-NPH53 | -0.8 | -4.7 | S-BSM2 | 1.0 | 6.3 | 2.17 |
| S-NPH3 | -0.6 | -3.8 | S-FTM16 | 1.0 | 6.0 | 2.26 |
| S-NPH4 | -0.6 | -3.6 | S-BSM9 | 0.8 | 5.0 | 2.28 |
| S-LAH71 | -0.9 | -5.5 | S-FPL51 | 1.5 | 9.2 | 2.51 |
| S-NPH1 | -1.3 | -8.2 | S-FTM16 | 1.8 | 11.0 | 2.60 |
| S-NPH2 | -0.5 | -3.0 | spinel | 0.6 | 3.7 | 2.87 |

## 4. DISTORTION

At conceptual design, we completed a preliminary study of distortion that suggested distortion levels were not objectionable, at least in the imager; however, it was recognized this was an area needing more study. In this phase we spent considerable time characterizing distortion. With the doubling of the field of view of the imager, distortion has become a major concern/risk factor. This may seem surprising, as the ADC is located in a collimated beam, but given the very stringent astrometric requirements, it should not be.

We find that ADC-induced distortion is composed of two elements, which we call "linear distortion" and "elevation distortion," described in the following figures. It is possible to characterize these terms with simple models, and such models will need to be applied for analysis of astrometry data. Also, we find that the magnitude of the distortion will limit exposure times to ~20 minutes (worst case) to avoid blurring the images due to *changing* distortion across the field!

Figure 2 shows representative distortion for a prism rotation angle $\theta = 45°$; this particular case is based on glass pair S-LAH79+S-FPM3. In this and the following figures, the green box represents the detector area, and the vector at right provides the scale of the distortion vectors. The data was produced by a specially written ray-tracing code that allowed easy exploration and testing of distortion models; the results of the code were checked against ZEMAX ray-tracing to confirm accuracy. We used this software to characterize the distortion models and to explore the dependencies of magnitude on prism rotation (or equivalently telescope elevation), glasses, prism tilts, scaling with distance from the field center, etc. Glass pairs studied were selected for a large range in parameters including absolute and relative refractive indices.

The distortion in Figure 2 is clearly complex. We find that it decomposes into the two patterns identified above, and we provide a description of each. The characteristics are summarized in Table 2.

Table 2. Characteristics of distortion terms with prism rotation and distance from field center.

| Distortion Type: | Scaling wrt $\theta$ | Scaling in field | relative amplitude |
|---|---|---|---|
| Elevation (y; x=0) | $\cos(\theta)$ | $r^2$ | glass |
| Linear-x | $\cos(\theta) \times \sin(\theta)$ | y | glass/tilts |
| Linear-y | $\cos(\theta) \times \sin(\theta)$ | x | glass/tilts |
| Pupil displacement | $\sin(\theta)$ | n/a | glass/tilts |

### 4.1 Elevation Distortion

This distortion is so named because the distortion vectors are always in the elevation direction. The magnitude varies as $\cos \theta$, and is thus greatest at maximum correction/lowest elevation. The magnitude also varies as $r^2$ away from the field center. Its amplitude depends strongly on the glasses in the pair, and is worst when the prism wedge angles (or glass indices) are significantly mismatched. It is not particularly sensitive to prism tilts. See Figure 3 for the appearance of elevation distortion in the absence of linear distortion. This distortion tends to dominate, and it appears to be intrinsic to the crossed-Amici prism design.

### 4.2 Linear Distortion

The magnitude of this distortion varies as $\sin \theta \times \cos \theta$ and is thus maximum at $\theta = 45°$. The magnitude varies as $r$ away from the field center. There is a very strong dependence of amplitude on prisms tilts; with careful alignment of the prism tilts, this distortion can be eliminated in principle. In Figure 4 and Table 3, the effect is calculated for a typical tolerance (i.e., error) in setting the prism tilt in a cell (that is, about 0.001-inch across the diameter of the prism). Linear distortion is thus extremely sensitive to mismatches in prism tilts, and properly tilting/aligning the prisms will be a major challenge. Fortunately, some linear distortion is correctable with NFIRAOS, although large corrections would quickly eat up the available stroke in the deformable mirrors.

When both distortion patterns are removed (see example in Figure 5), the residual distortion becomes the same order of magnitude as the astrometric precision required for the most stringent IRIS science cases. It is clear that distortion modeling will be a crucial component of the data reduction process, and images will need to be remapped before co-adding to remove the effects of a changing distortion pattern.

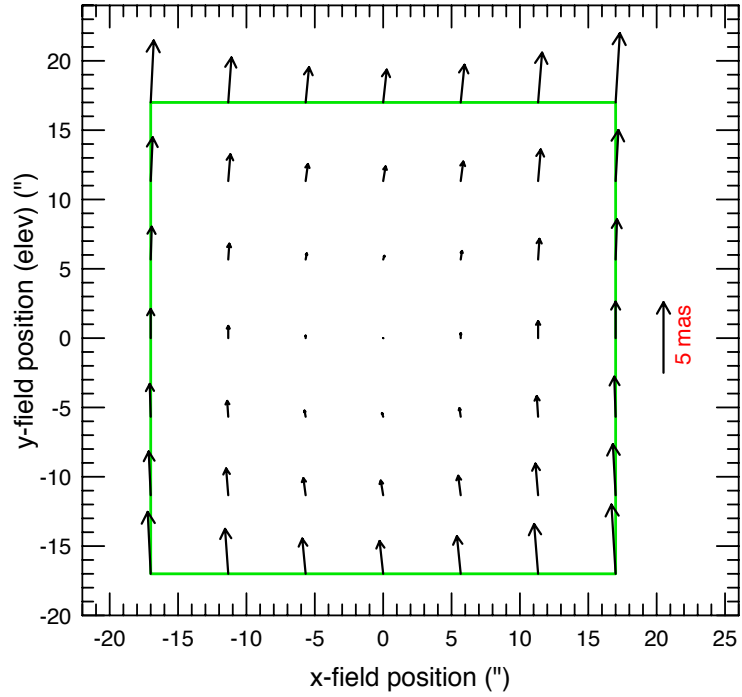

Figure 2. Calculated distortion for a representative glass pair, and prism rotation of 45° (corresponding to zenith distance ~57°). See text for more details.

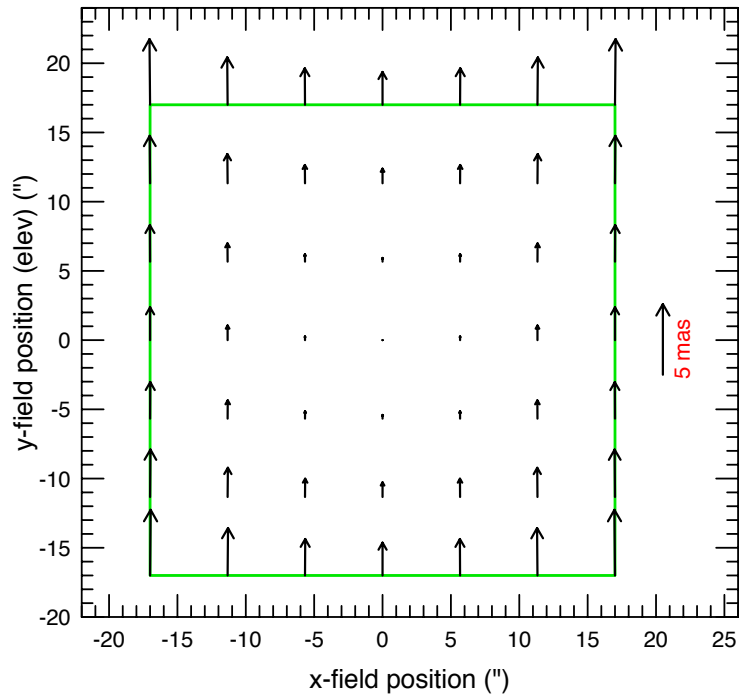

Figure 3. Same as Figure 2, except that a model of the "linear distortion" has been removed. The remaining pattern is representative of pure "elevation distortion." See text.

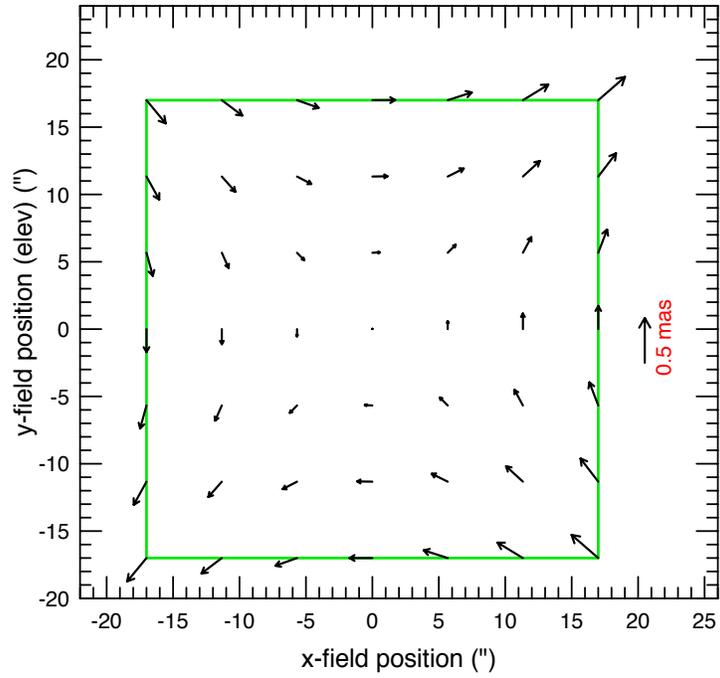

Figure 4. Same as Figure 3, except here a model of the elevation distortion has been removed, and the remaining pattern is representative of pure linear distortion. See text.

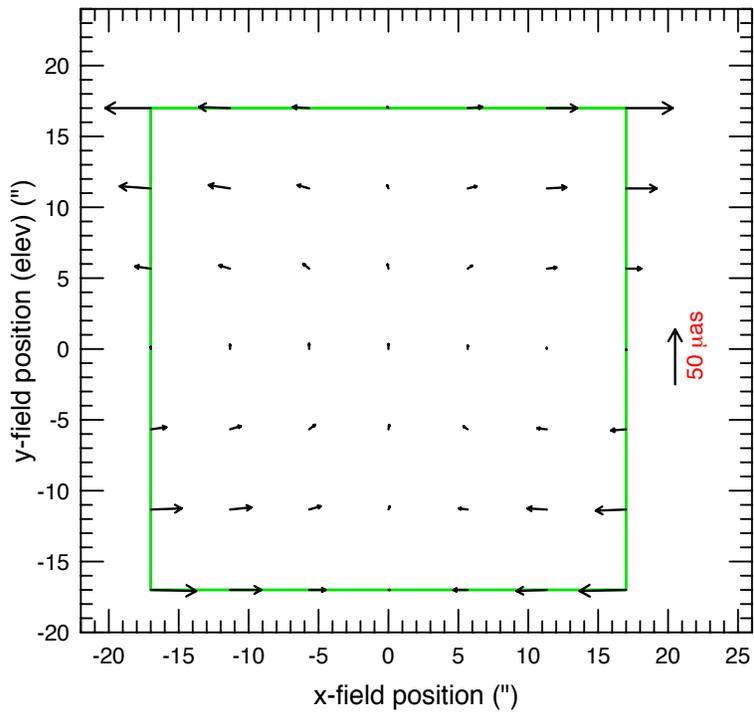

Figure 5. This shows the residual distortion in Figure 2 after both elevation and linear distortion models have been removed. The remaining residuals are now comparable to the astrometric precision required.

## 5. PUPIL DISPLACEMENT

The ADC in the collimated beam will generally have some effect on the pupil location as seen by the collimator. With the current fast TMA collimator design, pupil displacement can significantly affect image quality. The cause of this displacement is easily seen in Figure 6. In the case of maximum corrections (left) there is an over all symmetry, and any displacement introduced by the first Amici prism is removed by the second prism. However, in the nulled position (right), the prisms will act effectively as a tilted plate – actually two tilted plates at different angles. Unless the refractive indices and tilts are perfectly matched to cancel, there will be a net pupil shift. Given that the prism tilts must be carefully adjusted to minimize linear distortion, there is no leeway to adjust them to minimize pupil displacement. However, it appears that NFIRAOS can adjust for the effects of pupil shift at the levels we see it in the designs (see Table 3).

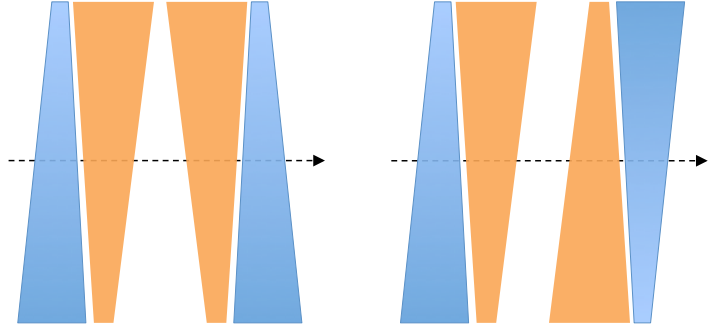

Figure 6. The ADC at full correction (side view, left) and nulled (top view, right), illustrating the source of pupil displacement. When at maximum correction, the symmetry means the ray will emerge at the same height as it entered. In contrast, when nulled, the prisms act as a tilted plate, which causes an offset in the exiting ray. This offset results in a displacement of the pupil.

## 6. FINAL DESIGN AND PERFORMANCE

We selected six pairs of glasses from Table 1 for further study; this are listed in Table 3. Our top choice was S-LAH79+S-FPM3, since it has very low residual dispersion and good transparency at λ > 2 μm. The others pairs were selected for various characteristics: S-LAH71+S-FPL51 for improved *K*-band throughput; S-TIM39+S-BSM2 for small residuals; S-FTM16+S-BAL42 for well-matched indices and prism angles; and S-TIH11+S-PHM52 and S-PBH56+S-BSM2 as having intermediate properties. We determined the precise wedge angles required, and found the prisms tilts that minimized linear distortion. We then calculated the distortion amplitudes, pupil displacement and expected combined transmission at λ > 2 μm with realistic thickness based on prism face slope. These values are all listed in Table 3. We then obtained quotes for the glass blanks. It turned out that two pairs, including our top choice, are currently unavailable in the required sizes.

Table 3: Summary of distortions and other parameters for six glass pairs under consideration for final choice

|  | S-LAH79 S-FPM3 | S-LAH71 S-FPL51 | S-TIM39 S-BSM2 | S-TIH11 S-PHM52 | S-PBH56 S-BSM2 | S-FTM16 S-BAL42 |
|---|---|---|---|---|---|---|
| Elevation Ampl. ($A_1$) | 1.056 | 1.010 | 0.256 | .379 | 0.567 | 0.020 |
| Linear Ampl. ($A_2$) | 0.60 | 0.77 | 1.2 | 0.67 | 0.66 | 1.37 |
| Pupil displacement (mm) | 0.88 | 1.02 | 1.61 | 0.89 | 0.88 | 1.87 |
| rms dispersion (mas) | 0.84 | 2.51 | 0.82 | 2.11 | 1.14 | 1.21 |
| Available? (D=140mm) | no | yes | yes | yes | no/yes | yes |
| Tot transmiss. @2.0 μm | 0.95 | 0.97 | 0.89 | 0.86 | .94 | 0.90 |
| Tot transmiss. @2.2 μm | 0.89 | 0.95 | 0.75 | 0.75 | .83 | 0.72 |
| Tot transmiss. @2.4 μm | 0.72 | 0.88 | 0.59 | 0.65 | .68 | 0.61 |

In the end, the decision was made that *K*-band throughput was most important to the IRIS science case and dominated the trade-offs. This restricted the choice to S-LAH71+S-FPL51. Even here, obtaining blanks of S-LAH71 of the required size proved problematic, so we had to move the ADC to just behind the pupil to accommodate a maximum blank diameter of 140 mm.

The final parameters for the ADC prisms are given in Table 4, and the residual dispersions are shown for the whole IRIS range as well as for individual passbands in Figure 7.

Work remaining on the optical design includes defining prism anti-reflection coatings, verifying that ghosting is acceptable, and designing an alignment plan.

Table 4. The parameters of the adopted ADC design

|  | **Prism 1** | **Prism 2** |
|---|---|---|
| Glass | S-LAH71 | S-FPL51 |
| Wedge angle (°) | 2.812 | -4.692 |
| Tilt (°) | 0 | 0.085 |
| Thickness (mm) | 12 mm | 12 mm |
| Gap (mm) | 4 mm | |
| max. Z full correction (for MK) | 65° | |

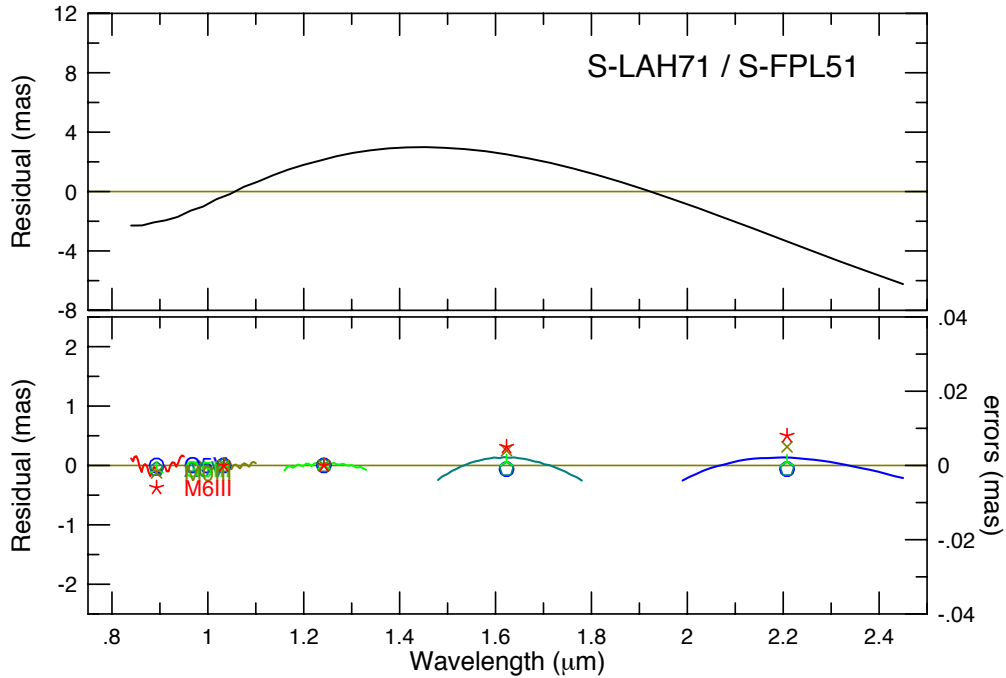

Figure 7. Residual dispersion for the selected glass pair, tuned to the entire wavelength range of IRIS (top) and to each individual passband (bottom). The scale in milli-arcseconds is shown on the left. Peak-to-valley residuals are well under 1 mas within each passband. The symbols in the lower panel represent the residuals weighted by five different stellar SEDs (O5V, A0V, G5III, M0III and M6III); they are plotted with the scale on the right. It is seen that these SED effects do not exceed 10 µas when properly tuned.

## ACKNOWLEDGMENTS

The TMT Project gratefully acknowledges the support of the TMT collaborating institutions. They are the California Institute of Technology, the University of California, the National Astronomical Observatory of Japan, the National Astronomical Observatories of China and their consortium partners, the Department of Science and Technology of India and their supported institutes, and the National Research Council of Canada. This work was supported as well by the Gordon and Betty Moore Foundation, the Canada Foundation for Innovation, the Ontario Ministry of Research and Innovation, the Natural Sciences and Engineering Research Council of Canada, the British Columbia Knowledge Development Fund, the Association of Canadian Universities for Research in Astronomy (ACURA), the Association of Universities for Research in Astronomy (AURA), the U.S. National Science Foundation, the National Institutes of Natural Sciences of Japan, and the Department of Atomic Energy of India.

## REFERENCES

[1] Larkin, J. E., Moore, A. M., Wright, S. A., Wincentsen, J. E., Chisholm, E. M., Dekany, R. G., Dunn, J. S., Ellerbroek, B. L., Hayano, Y., Phillips, A. C., Simard, L., Smith, R., Suzuki, R., Weiss, J. L. and Zhangh, K., "The Infrared Imaging Spectrograph (IRIS) for TMT: instrument overview," Proc. SPIE 9908, in press (2016).
[2] Herriot, G., Andersen, D., Atwood, J., Boyer, C., Byrnes, P., Caputa, K., Ellerbroek, B., Gilles, L., Hill, A., Ljusic, Z., Pazder, J., Rosensteiner, M., Smith, M., Spano, P., Szeto, K., Véran, J-P. Wevers, I., Wang, L., Wooff, R., "NFIRAOS: first facility AO system for the Thirty Meter Telescope," Proc. SPIE 9148, 914810 (2014).
[3] Phillips, A. C., Bauman, B. J., Larkin, J. E., Moore, A. M., Niehaus, C. N., Crampton, D. and Simard, L., "The Infrared Imaging Spectrograph (IRIS) for TMT: the atmospheric dispersion corrector," Proc. SPIE 7735, 77355Q (2010).
[4] Dunn, J., Reshetov, V., Atwood, J., Pazder, J., Wooff, B., Loop, D., Saddlemyer, L., Moore, A. M. and Larkin, J. E., "On-instrument wavefront sensor design for the TMT infrared imaging spectrograph (IRIS) update," Proc. SPIE 9147, 91479H (2014).